# Delineating Geographical Regions with Networks of Human Interactions in an Extensive Set of Countries

**Stanislav Sobolevsky[1]\*, Michael Szell[1], Riccardo Campari[1], Thomas Couronné[2], Zbigniew Smoreda[2], Carlo Ratti[1]**

1 Senseable City Laboratory, Massachusetts Institute of Technology, Cambridge, Massachusetts, United States of America, 2 Sociology and Economics of Networks and Services Department, Orange Labs, Paris, France

## Abstract

Large-scale networks of human interaction, in particular country-wide telephone call networks, can be used to redraw geographical maps by applying algorithms of topological community detection. The geographic projections of the emerging areas in a few recent studies on single regions have been suggested to share two distinct properties: first, they are cohesive, and second, they tend to closely follow socio-economic boundaries and are similar to existing political regions in size and number. Here we use an extended set of countries and clustering indices to quantify overlaps, providing ample additional evidence for these observations using phone data from countries of various scales across Europe, Asia, and Africa: France, the UK, Italy, Belgium, Portugal, Saudi Arabia, and Ivory Coast. In our analysis we use the known approach of partitioning country-wide networks, and an additional iterative partitioning of each of the first level communities into sub-communities, revealing that cohesiveness and matching of official regions can also be observed on a second level if spatial resolution of the data is high enough. The method has possible policy implications on the definition of the borderlines and sizes of administrative regions.

**Citation:** Sobolevsky S, Szell M, Campari R, Couronné T, Smoreda Z, et al. (2013) Delineating Geographical Regions with Networks of Human Interactions in an Extensive Set of Countries. PLoS ONE 8(12): e81707. doi:10.1371/journal.pone.0081707

**Editor:** Yamir Moreno, University of Zaragoza, Spain

**Received** August 23, 2013; **Accepted** October 25, 2013; **Published** December 18, 2013

**Copyright:** © 2013 Sobolevsky et al. This is an open-access article distributed under the terms of the Creative Commons Attribution License, which permits unrestricted use, distribution, and reproduction in any medium, provided the original author and source are credited.

**Funding:** Financial supporters of the Senseable City Laboratory are: the National Science Foundation, the MIT SMART program, the Center for Complex Engineering Systems (CCES) at KACST and MIT, AudiVolkswagen, BBVA, The Coca Cola Company, Ericsson, Expo 2015, Ferrovial and all the members of the MIT Senseable City Lab Consortium. The funders had no role in study design, data collection and analysis, decision to publish, or preparation of the manuscript.

**Competing Interests:** We have the following interests: This study was partly funded by AudiVolkswagen, BBVA, The Coca Cola Company, Ericsson, Expo 2015 and Ferrovial. Orange, British Telecom, Telecom Italia and Saudi Telecom Company provided datasets for this research. This does not alter the authors' adherence to all the PLOS ONE policies on sharing data and materials.

\* E-mail: stanly@mit.edu

## Introduction

In recent years, human geography and many other areas of social science have been experiencing exciting new developments due to the availability of large-scale data from human interactions, communications, and movements. Advances in information and communication technologies, and the accumulation of massive data sets on human behavior now allow researchers to study human interaction and mobility patterns with unprecedented precision [1]. Data from human interactions such as mobile phone usage can provide insights on various questions in human geography which otherwise would be impossible to understand quantitatively. The issues that can now be tackled in unprecedented details concern fields as diverse as geomarketing [2], urban planning [3], having implications for epidemiology and spread of diseases [4] or generally for the spread of information [5] and the understanding of individual mobility patterns [6] and political movements [7]. Even purely virtual environments have the potential to advance our understanding of the nature of human movements [8].

Here our approach focuses on the topology of human interactions in the form of networks. A network viewpoint emphasizes that the behavior of a complex system is shaped by the strong interactions among its constituents and offers the possibility to analyze social systems within an abstracted, mathematically well-tractable framework [9]. Our main point of interest is the partitioning of human population in space based on the raw networks of communication activities. We build on a small corpus of previous studies [10–13], in which human activity networks within single countries have been studied. The partitioned networks have shown to reflect the linguistic or cultural borders of underlying geographical space, and to follow administrative boundaries, sometimes surprisingly close. We now bring together large data sets from a number of different countries, broadening the scope to a multitude of regions and cultural backgrounds, showing that the observed effects tend to hold in general, and also on a second level of partitioning if the given data is fine-grained enough to allow such a partitioning. To this end we employ community detection algorithms which optimize modularity as in previous works. Further, comparing proposed borders to the underlying regions given by the human activity data in a rejected administrative referendum in Portugal demonstrates the practical implications of our work, able to reveal the actual underlying social structure of the population and to provide "ground truth" to decision makers.

Community detection of phone call networks via modularity optimization, see Materials and Methods, was established in previous works [10,11], leading to spatially cohesive regions generally consistent with the geopartitioning of major political regions of the considered countries. Communication networks have been shown to be a reasonable proxy for other human interaction networks [14,15], making the observations generalizable to human activity beyond phone calls. A massive communication network inferred from a large telecommunications database in Great Britain based on landline calls has been studied previously [10]. The study found geographically cohesive regions that generally correspond with administrative regions, while unveiling unexpected spatial structures that had previously only been hypothesized in the literature. The cohesiveness of single regions was assessed [10], showing how "integrated" different regions of the UK are. Further, communication networks of Belgium based both on average duration of calls and on the total call duration were analyzed [11]. The latter study has yielded a number of groups which the authors called *spatially balanced*: the 17 areas found resemble the urban hierarchy suggested earlier [16]. Further, these groups are always made up of *adjacent* municipalities although this is not a necessary outcome of the algorithm. Both results were also found for France [12]. Finally, the partitioning of an average call duration network in Belgium delineated exactly two areas which closely follow the linguistic border inside the country. Language therefore seems to constitute a strong barrier in human spatial organization from communication.

Further related work has partitioned data of commutes in the US and has also found regions being both cohesive and following borders [13]. Related is a work which analyzed the movements of virtual avatars in a massive multiplayer online game [8]. In this case the detection of communities from networks of raw mobility data has yielded an almost exact match with underlying socioeconomic regions of the virtual society. The study also quantified the strong influence of borders on mobility. Similar investigations on borders were performed using networks of money flows [17], or using GPS tracks of vehicles and an Infomap approach to compare detected clusters with existing administrative borders on a more local level [18]. Mobility of mobile phone users has been explored for the country of Portugal [19].

## Materials and Methods

### Data sets

We consider seven country-wide data sets of telephone calls, in France, UK, Italy, Belgium, Portugal, Saudi Arabia, and the Ivory Coast, details on the data is given in Table 1. All data sets comprise mobile phone data with the exception of landline calls in UK and Italy. Data was provided by single market providers with possible heterogeneous coverage over the respective countries – we have no information on local market shares and on resulting possible inhomogeneities in spatial coverage. The Ivory Coast data was released to researchers during the D4D mobile phone data challenge [20] and was used as is. All other data sets are proprietary and subject to stricter data privacy agreements, therefore here we do not have the possibility to provide more expressive information on metadata or on the data collection process available than provided in Table 1. All data has been anonymized and aggregated on the operator side prior to receipt and in line with all local data protection laws. There was no special cleaning process performed which could have introduced substantial bias. All the operators who provided the data possess country-wide coverage for the corresponding countries. In those cases where the coverage ratio was substantially spatially inhomogeneous the appropriate normalization by the local market shares has been performed for the aggregated communication networks.

## Network extraction

We construct interaction networks between different locations of a country based on the aggregated duration of calls having origin in the first and destination in the second location. This process generates a weighted directed network in which the loop edges from locations to themselves are also considered. We construct the aggregated networks of communication flows between all given different locations of the country at the available spatial resolution level defining a link weight between each two locations as a total duration of calls initiated by the users of the first considered location to the users of the second one. The nodes in these networks are the locations, ranging from municipalities, zip codes, special geographical units such as exchange areas, or cell tower areas, as defined in the "Spatial resolution" column of Table 1. In case of Portugal and France the users are attached to their actual locations during a call, while for Belgium to their formal residence locations. The UK and Italy networks are based on landline calls, i.e. the locations of the users are fixed.

## Partitioning algorithm

To the extracted communication networks we apply an algorithm for community detection following a standard modularity optimization approach [21,22]. The method scores all the edges of the network according to their relative strength compared to a null-model with respect to the weight of the nodes they connect and aims to maximize the cumulative score inside the communities, preferring edges with a positive score and avoiding those with a negative score. The particular optimization algorithm [23] is a variation of the technique used by [10]. The idea is an iterative improvement of the partitioning in terms of the modularity score, starting from a trivial case where all nodes are gathered into one community involving three kinds of possible improvements: 1) dividing a community into two new communities, 2) joining two communities into one, and 3) shifting a part of one community to another existing community. The outcome of partitioning spatial networks is in general not qualitatively dependent on the particular algorithm used – the reason we use this one is because of its ability to consistently provide the best results in terms of modularity score compared to other algorithms, [23], see also SI.

The intriguing property of the modularity optimization approach is that the resulting network division has no predetermined number of partitions. Only the raw topological information of the input network determines the range of communities detected. Further, the algorithm does not fix the sizes nor the distribution of sizes of the detected groups, and it is not limited by any spatial constraints.

Boundaries produced by the algorithm which match official boundaries might naively be interpreted as having a "natural" validation of the hypothesis of closely followed borders – two divisions of a country, partitions from networks and official borders, would not coincide just by chance but rather for a reason. However, if the algorithm's result does *not* match, the reasons – apart from a genuine deviation of human interaction regions from official boundaries – could also include low population density near the border making boundaries visually floating but leaving modularity scores practically unchanged, and other possible minor statistical fluctuations. Due to such influences, the boundaries produced by the algorithm cannot always be treated as exact, they may be shifted slightly. However, the cores of detected regions have found to be stable for the UK data set [10].

**Table 1.** Properties of the data sets.

| Data set | Type | Dur. | Calls | Phones | Time | Spatial resolution $n$ | Dir. |
|---|---|---|---|---|---|---|---|
| France | Mobile | 120 bn. | 800 m. | 14 m. | 45 days | 17,800 cell towers | yes |
| UK | Landline | 452 bn. | 7.6 bn. | 47 m. | 1 month | 4800 exchange areas | no |
| Italy | Landline | 410 bn. | n/a | n/a | 55 days | 200 regions | yes |
| Belgium | Mobile | 35 bn. | 200 m. | 2.6 m. | 6 months | 600 municipalities | yes |
| Portugal | Mobile | 56 bn. | 440 m. | 1.6 m. | 15 months | 2200 cell towers | yes |
| Ivory Coast | Mobile | 6 bn. | 62 m. | 5 m. | 2 weeks | 1100 cell towers | yes |
| Saudi Arabia | Mobile | 570 bn. | 2.7 bn. | 14 m. | 30 days | 500 cell towers | yes |

Country-wide telephone data sets are provided by single telephone operators, covering different time frames, with different numbers of phones, calls, total call durations (Dur.) and on various spatial resolutions. The abbreviations bn. and m. stand for billion and million, respectively. Resolution numbers are given as approximate values. These locations constitute the nodes of the corresponding telephone call networks, while the sum of durations of calls between locations span their weighted links. The last column (Dir.) denotes if the network is directed.
doi:10.1371/journal.pone.0081707.t001

## Partition overlap measures

We use two classical measures of clustering similarity to quantify partition overlap, i.e. of how well two different partitions of the same set of locations match: Rand's criterion $\mathcal{R}$ [24] and the Fowlkes and Mallows index $\mathcal{F}$ [25]. Both of these measures are based on comparing sets of pairs of locations which have either the same community in both partitions or a different community. A perfect match between two partitions will have $\mathcal{R}, \mathcal{F} = 1$. For the case of two completely unrelated clusterings, both indices are in general strictly larger than zero, more so for $\mathcal{R}$ [25]. Therefore, to have a baseline, we calculated the average indices over 1000 random reshufflings of locations in given administrative regions, denoted by $\mathcal{R}_r$ and $\mathcal{F}_r$. To have a measure grounded in another, information-theoretical approach, we also use the variation of information $VI$. The $VI$ has mathematical properties that are in line with our general intuition of what "more different" and "less different" should mean for two clusterings of a set [26]. For formal definitions of all measures see SI.

## Results

### Confirmation of cohesiveness and border-similarity in an extended set of countries

From the few previously studied cases, we considerably extend the number of countries in which we apply the partitioning algorithm to country-wide phone communication networks, involving the large-scale European countries France, UK, Italy, the smaller countries of Portugal and Belgium, as well as the African and Asian countries of Ivory Coast and Saudi Arabia. Figures 1, 2, and 3, and Tables 2 and 3 show all results and key statistics for the considered countries. On the one hand black borders in the figures show the official administrative regions. For UK, we display the level 1 NUTS regions of the European Union defined by the nomenclature of territorial units for statistics (http://ec.europa.eu/eurostat/ramon), for all other European countries the level 2 NUTS regions (NUTS2), for Ivory Coast the countries' regions, for Saudi Arabia its provinces. For Portugal we additionally show historical regions in Fig. 2D and the borders of a proposed, failed referendum of 1998 in Fig. 2E. On the other hand colored areas display the partitioning from our community detection algorithm applied to the communication networks.

We first reproduce the partitioning for UK and Belgium from previous works [10,11], see Fig. 1C and Fig. 2A respectively. As previously found, the resulting detected communities are geographically cohesive regions following to most parts official borders of the countries. Quantified by the clustering indices, the UK partitioning shows values of $\mathcal{R} = 0.955$ with a baseline of $\mathcal{R}_r = 0.809$ and $\mathcal{F} = 0.772$ with a baseline of $\mathcal{F}_r = 0.107$, while Belgium has $\mathcal{R} = 0.932$ with a baseline of $\mathcal{R}_r = 0.819$ and $\mathcal{F} = 0.647$ with a baseline of $\mathcal{F}_r = 0.101$. Due to slight differences in the data sets and the higher efficiency of the algorithm used, our partitionings show small deviations to the previous ones [10,11]. For example, in Belgium, the previously delineated separate parts of West and East Flanders are now merged together into connected regions. We report the result for France in Fig. 1A. Almost all borders are followed strikingly well, with only two exceptions: The regions of Limousine and Auvergne are to most parts joined together, Rhône-Alps is split into three, were the southern part belongs to an area which encompasses Languedoc-Roussillon and a western piece of Provence-Alpes-Côte d'Azur. The clustering indices are correspondingly high: $\mathcal{R} = 0.985$ with a baseline of $\mathcal{R}_r = 0.860$ and $\mathcal{F} = 0.900$ with a baseline of $\mathcal{F}_r = 0.076$.

Figure 1E and Fig. 2C show the results for the newly considered European countries Italy and Portugal, respectively. Again we detect only cohesive regions, and again these regions follow political boundaries closely. The clustering indices for Italy read $\mathcal{R} = 0.957$ with a baseline of $\mathcal{R}_r = 0.883$ and $\mathcal{F} = 0.647$ with a baseline of $\mathcal{F}_r = 0.063$, for Portugal $\mathcal{R} = 0.885$ with a baseline of $\mathcal{R}_r = 0.677$ and $\mathcal{F} = 0.697$ with a baseline of $\mathcal{F}_r = 0.203$. Visually, most boundaries in Portugal follow the borders of the political NUTS2 provinces well, however some of them are merged together like Baixo Alntejo and Algarve (the two most southern provinces) as well as all the Beira provinces (Beira Litoral, Beira Alta, Beira Baixa) of the bigger historical region of Centro. Because of this, we consider additionally to the NUTS2 borders the similar but more fine-grained borders of 11 historical provinces, going back to the Administrative Code of 1936 [27], Fig. 2D. These historical borders can explain some of the deviations, supported by the improved clustering indices $\mathcal{R} = 0.919$ and $\mathcal{F} = 0.742$, however they also introduce a surplus of internal boundaries.

The partitioning of Italy generally corresponds to the official NUTS2 division, but also here a number of border shifts between neighboring regions are observed. Some notable deviations from official borders include the city of Verona being part of the region Trentino-Alto Adige, and the most eastern part of Liguria, La spezia, being joined together with the region found for Tuscany. We also find additional, split up regions, such as an additional small region in the western part of Emilia-Romagna resembling the historical Ducato di Parma e Piacenza (Duchy of Parma and Piacenza), or Sicily being split into three.

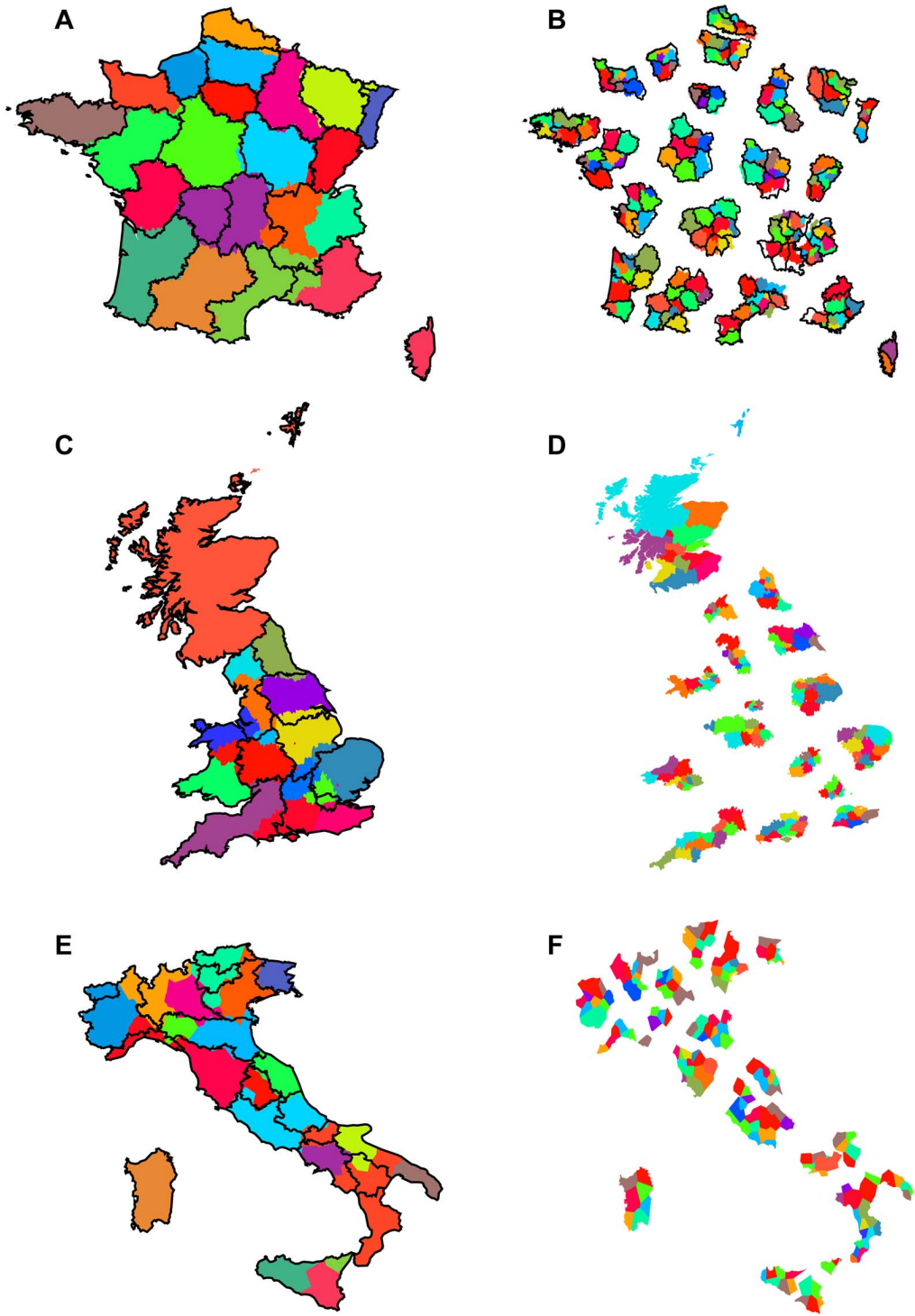

**Figure 1. Partitioning of large European countries based on telephone call networks.** Left column: Community detection (first level) of telephone call networks of (A) France, (C) UK, (E) Italy. The black lines show the 22, 11, and 20 administrative regions (NUTS1 for UK, NUTS2 for the other countries), respectively, the colored areas show the corresponding 21, 16, 22 level 1 regions found by applying the modularity optimization algorithm on the country-wide phone call networks. All detected regions are cohesive although some of the distinct colors used may appear similar. Right column: Community detection (second level) within all network partitions from the first level, of (B) France, (D) UK, (F) Italy. For visual clarity here we present the second level communities grouped into first level communities in an exploded view. Colors of detected subregions only apply inside their respective level 1 partitions, again all detected subregions are cohesive although some of the distinct colors used may appear similar. For France we also show the official NUTS2 borders which considerably match well the second level partitioning.
doi:10.1371/journal.pone.0081707.g001

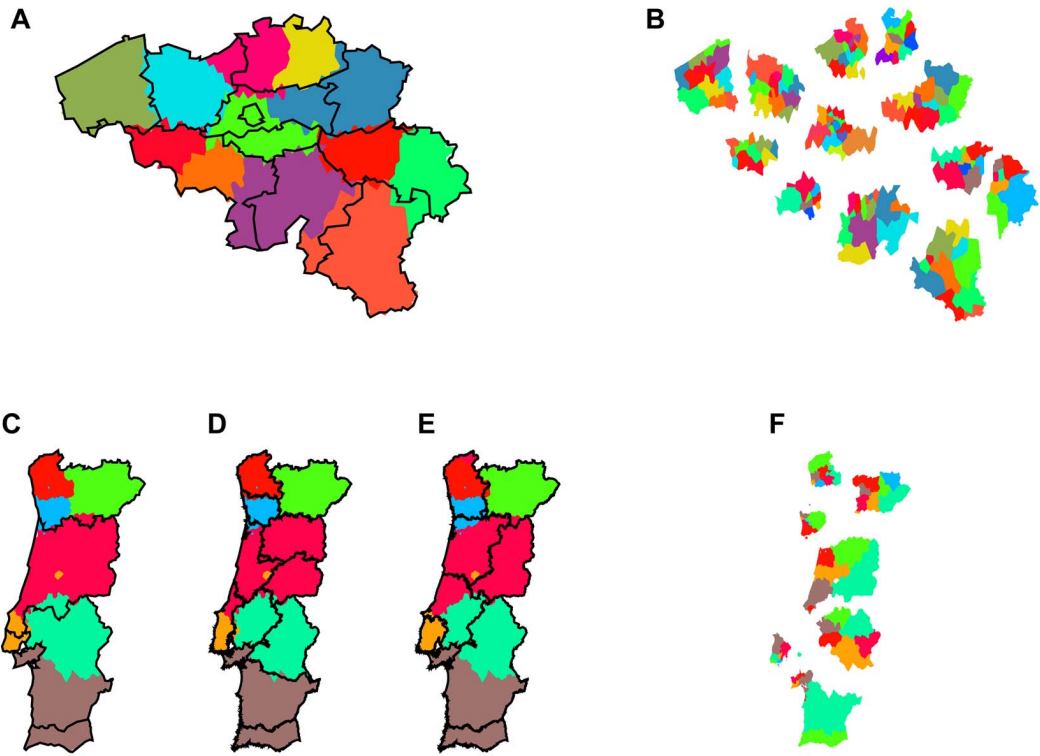

**Figure 2. Partitioning of small European countries based on telephone call networks.** Left column: Community detection (first level) of mobile phone call networks of (A) Belgium, (C) Portugal. The black lines show the 11 and 5 administrative regions (NUTS2), respectively, the colored areas show the corresponding 12, 7 level 1 regions found by applying the modularity optimization algorithm on the country-wide phone call networks. Panel (D) shows the 11 historical regions of Portugal, which can explain some of the deviations from NUTS2. Panel (E) shows the poorly matching borders proposed in the failed referendum of 1998 to restructure administrative regions. All detected regions are cohesive although some of the distinct colors used may appear similar. Right column: Community detection (second level) within all network partitions from the first level, of (B) Belgium, (F) Portugal. For visual clarity here we present the second level communities grouped into first level communities in an exploded view. Colors of detected subregions only apply inside their respective level 1 partitions, again all detected subregions are cohesive although some of the distinct colors used may appear similar.
doi:10.1371/journal.pone.0081707.g002

In the cases of Ivory Coast and Saudi Arabia, especially in the latter one, the matching is visually less clear, Fig. 3. This observation is also expressed in the relatively low cluster overlap indices: $\mathcal{R} = 0.870$ with a baseline of $\mathcal{R}_r = 0.739$ and $\mathcal{F} = 0.505$ with a baseline of $\mathcal{F}_r = 0.154$ for Ivory Coast, and $\mathcal{R} = 0.904$ with a baseline of $\mathcal{R}_r = 0.795$ and $\mathcal{F} = 0.606$ with a baseline of $\mathcal{F}_r = 0.116$ for Saudi Arabia. The $VI$ index is slightly above 2 for both countries, which is not the case for other countries. We assume this relatively poorer agreement stems from the vast spatial

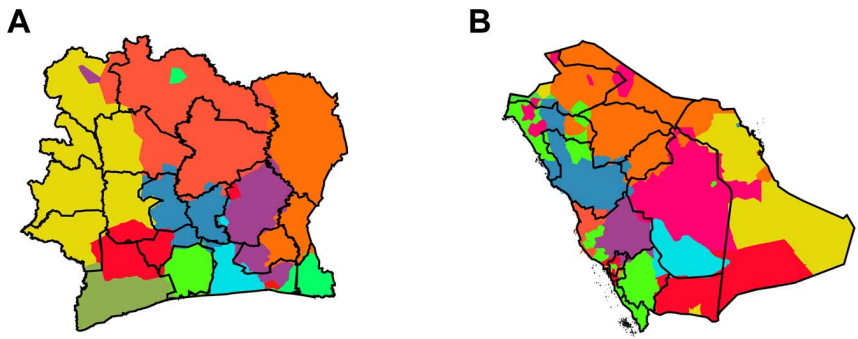

**Figure 3. Partitioning of countries based on telephone call networks.** (A) Ivory Coast and (B) Saudi Arabia. The black lines show the 19 and 13 administrative regions, respectively, the colored areas show the corresponding 11 and 12 level 1 regions found by applying the modularity optimization algorithm on the country-wide phone call networks. For these two countries, the matching shows less overlap than for other countries, likely due to the low population density and the sparse data. Almost all detected subregions are cohesive, exceptions appear in Saudi Arabia which is very heterogeneously populated. Due to the heterogeneous distribution of cell towers, a second level partitioning is not feasible here.
doi:10.1371/journal.pone.0081707.g003

**Table 2.** Basic statistics for countries with number of various regions.

| Country | Pop. | Dens. | NUTS1 | NUTS2 | NUTS3 | Alt. | L1 | L2 |
|---------|------|-------|-------|-------|-------|------|-----|-----|
| France* | 65 | 118 | 8 | 22 | 96 | – | 21 | 207 |
| UK** | 63 | 257 | 11 | 36 | 128 | – | 16 | 150 |
| Italy | 61 | 206 | 5 | 21 | 107 | – | 21 | 150 |
| Belgium | 11 | 359 | 3 | 11 | 44 | – | 12 | 129 |
| Portugal*** | 11 | 116 | 1 | 5 | 28 | 11 | 7 | 44 |
| Ivory Coast | 21 | 64 | – | – | – | 19 | 11 | – |
| Saudi Arabia | 29 | 12 | – | – | – | 13 | 12 | – |

All columns except Population and Density limited to:
*Metropolitan France,
**Britain,
***Mainland.
Density (Dens.) is given in population per square kilometer, population data (Pop.), in millions, is taken from the World Bank, 2011 (http://data.worldbank.org). The columns NUTS1, NUTS2, and NUTS3 refer to the numbers of official administrative regions. Alternative partitions (Alt.) refer to the 11 historical regions of Portugal going back to the Administrative Code of 1936 [27], to the 19 regions of Ivory Coast, and to the 13 regions of Saudi Arabia. Column L1 shows the number of communities found by the community detection algorithm, L2 the number of sub-communitites after iterative application to subnetworks.
doi:10.1371/journal.pone.0081707.t002

extent of these countries combined with the low population density, Table 2, and their more inhomogeneous distribution of cell phone towers.

## Iterative partitioning of subregions reveals similar properties on a second level

By partitioning the country-wide networks of human telephone interactions we obtain spatially cohesive regions generally consistent with the geopartitioning of greater political regions. However, it is possible to go one step further to apply the community detection method in an iterated fashion. Namely, applying the network partitioning to the subnetwork inside each of the detected first-level regions allows to produce a second-level subpartitioning of the network into smaller subregions. The panels in the right column of Figs. 1 and 2 show that second-level subpartitioning again possess the same general properties – all the subregions are geographically cohesive.

Since the match of first level regions with NUTS2 is most consistent for France, this makes it possible to also compare level two regions with NUTS3 in this country without running into too many inconsistencies due to deviations on the first level. These inconsistencies result in visual artifacts, see the few "hollow" regions in Fig. 1B. Although the number of level two regions found (207) is higher than the number of existing NUTS3 regions (96 without the overseas department), many borders are again followed reasonably well. The most visible mismatches occur in the same south-eastern parts were already level 1 regions are mismatched.

## Deviations

The findings of our approach, especially deviations from specific borderlines, have the additional potential to serve as decision aid for administrative officials and regional planners, either for or against specific possible subdivisions of a country, as well as give an insight into the long-time geographic impacts of historic events. Using telephone call data, which is recorded and stored by telephone providers and is therefore relatively easy to access by the respective companies, has the added benefit of being several orders of magnitudes less costly than performing censuses.

**Table 3.** Overlap indices for different countries.

| Country | $\mathcal{R}_\varepsilon$ | $\mathcal{R}$ | $\mathcal{F}_\varepsilon$ | $\mathcal{F}$ | $\log_2 n$ | $VI$ | Mod. |
|---------|------|------|------|------|------|------|------|
| France* | 0.860 | 0.985 | 0.076 | 0.900 | 14.12 | 0.676 | 0.78 |
| UK** | 0.809 | 0.955 | 0.107 | 0.772 | 12.21 | 1.322 | 0.62 |
| Italy | 0.883 | 0.957 | 0.063 | 0.647 | 7.79 | 1.349 | 0.72 |
| Belgium | 0.819 | 0.932 | 0.101 | 0.647 | 9.20 | 1.538 | 0.74 |
| Portugal*** | 0.677 | 0.885 | 0.203 | 0.697 | 11.08 | 1.465 | 0.49 |
| Ivory Coast | 0.739 | 0.870 | 0.154 | 0.505 | 10.12 | 2.054 | 0.37 |
| Saudi Arabia | 0.795 | 0.904 | 0.116 | 0.606 | 8.98 | 2.036 | 0.48 |

All columns limited to:
*Metropolitan France,
**Britain,
***Mainland.
$\mathcal{R}_\varepsilon$ and $\mathcal{F}_\varepsilon$ give the baselines for Rand's criterion $\mathcal{R}$ and the Fowlkes-Mallows index $\mathcal{F}$. The closer $\mathcal{R}$ and $\mathcal{F}$ to 1, the better the overlap of the detected communities with the administrative regions. On the other hand, $\log_2 n$ gives the upper limit for the $VI$ measure. Here, the closer $VI$ is to 0, the better the overlap. Modularity scores are shown in the last column (Mod.).
doi:10.1371/journal.pone.0081707.t003

In the following we highlight the case of Portugal and the administrative referendum of 1998 [28]. This referendum failed, as the majority of citizens voted against the newly proposed borders, shown in Fig. 2E. The proposed regions show a poor match with the regions from human interaction networks, also reflected in the clustering indices of $\mathcal{R} = 0.906$ and $\mathcal{F} = 0.714$ lying below the indices of the historical regions of $\mathcal{R} = 0.919$ and $\mathcal{F} = 0.742$ (but still above the NUTS2 values of $\mathcal{R} = 0.885$ and $\mathcal{F} = 0.697$), which is one of the possible reasons why the referendum failed. Comparing our partitioning result of Portugal to the today existing official territorial division of NUTS reveals that NUTS2 is more coarse-grained (5 regions in continental Portugal) while NUTS3 is more fine-grained (28 regions in continental Portugal) than our partitioning which is in-between (7 regions) and matches historical regions to some extent better. For example, the referendum proposed to split up Beira and to shift the Beira–Ribatejo border while not placing borderlines between i) Minho and Douro Litoral and between ii) Alto Alentejo and Baixo Alentejo, although these borders appear in the partition. These observations and the evidence from the clustering indices show that historical effects of human behavior could outlast modern categorization and might have an impact on policies today.

Another example of possible policy implications comes from the granularity of the results. While on the country-wide level the partitioning algorithm gives the closest match for NUTS2 regions in Belgium, Italy, and France, in the UK instead the same scale partitioning appears to match rather the NUTS1 definition. The definition of different levels of NUTS regions is known to be country-dependent. Therefore the deviation of the scale of NUTS regions in the UK from other EU countries may provide valuable input for creating a possibly more adequate definition of hierarchical regions with the aim to be homogeneous EU-wide, especially considering that different levels of NUTS regions correspond to very specific levels of structural fundings possibly impacting regional performance substantially [29].

For the case of Ivory Coast, the stronger deviations from political regions than in European countries possibly hints on one hand towards the mentioned inhomogeneous distribution of population or cell towers. On the other hand, the deviations might stem from the young age of the country's administrative structure, still going through processes of social reorganization after two recent civil wars. Here the present political borders, which are not fully consistent with earlier tribal structures, have been defined only a few decades ago as opposed to the long history behind the subdivision of France.

In conclusion, the regional structures based on the actual human interactions can be determined in an automated way independent of a country's history and can provide possible alternatives to existing administrative regions for organizing societies.

## Finding "breaking lines"

So far we used the partitioning program without any restrictions on number of communities and left it to the algorithm to find the most "natural" number in terms of modularity. In this section, we modify the algorithm which we have used above to limit this number to the smallest nontrivial number of communities, namely two, revealing the "breaking lines" of countries, i.e. the borders which split up countries into exactly two parts in terms of total network weight by optimizing modularity. In this case the algorithm optimizes the modularity value of all possible bisections, by restricting the ability of the algorithm to create a new community once the maximal allowed number of communities (here two) is already reached.

We report the results in Fig. 4. France is split by a border going from center north to center south following the eastern borders of the regions Upper Normandy, Île-de-France, Centre, Limousin, Midi-Pyrénées, Fig. 4A. The UK splits along a west-east line which also splits Wales in two, following the same split already found in level 1 regions, Fig. 4B. Mainland Italy is split along a line roughly following the northern border of Emilia-Romagna, Fig. 4C, Belgium is split along the Dutch-French language barrier with Brussels assigned to the northern Dutch part, Fig. 4D, Portugal is split slightly south of the Mondego river, Fig. 4E. A similar bi-split of Belgium was previously found [11], however it required a different measure for network links – average duration of one call – while the split we report here is obtained based on the same network of total call durations, but just with a limitation on the number of communities.

It is possible to quantify the strength of the splits by looking at the weights of links within each side compared to the total weight of the network. In all cases, we observe that the countries are divided into two parts with nearly equal network weight. Therefore, if the links were to be distributed homogeneously, we would expect around 50% of the links between the two split parts. However, the actual picture is quite different. Belgium displays the strongest split among all European countries with only 3.5% of all links going between the north and the south partition. The next strongest splits are France with 5.7% and Italy with 7.8% of links going between the split parts. The weakest splits are UK and Portugal with values of 9.5% and 12.1%, respectively. The modularity scores for the two-part partitions follow the same order and are: Belgium 0.46, France 0.44, Italy 0.42, UK 0.40, Portugal 0.38.

Note that the "breaking lines" and their strengths do not necessarily come with any political implications. First, a consequence of the algorithm is the separation of the network into two parts with almost equal total link weight. If we assume homogeneous communication behavior, then the population is expected to be divided in half by the process. Therefore, these splits have to be discussed with care, as results can be strongly influenced by heterogeneous population densities and/or geographically distinct features such as mountains. In some cases however, additional cultural reasons may be well justifiable. The most clear division in the case of Belgium falls in line with previous results where a strong language barrier was found between the northern Flemish region and the southern Walloon region [11]. In our case the bilingual city of Brussels is assigned to the northern instead of the southern partition, but it is not clear if this is simply due to the population distribution. Nevertheless, approximately half of the total 3.5% of links between the north and south areas go between Brussels and the south, making the capital a bridge between the regions and providing motivation for future studies on the human interactions within the Brussels region. Apart from clear cultural differences as in Belgium, the results in Italy might be influenced by migration patterns. Here, surprisingly we find that the western and southern islands of Sardinia and Sicily are connected to the northern partition. This could be due to the nature of the algorithm, which would assign possibly weak connections between the islands and the mainland less clearly. On the other hand, substantial migration flows from southern parts of Italy to the north since after the second world war are well known. Especially the north-western regions of Piedmont, Lombardy, Liguria and Aosta Valley were the destinations of a large proportion of the migration flows of the 1950s and 1960s, since industrial development in Italy has its origins there [30]. The connection of Sicily and Sardinia to this northern part could be due to family ties spanning between migrated and remaining

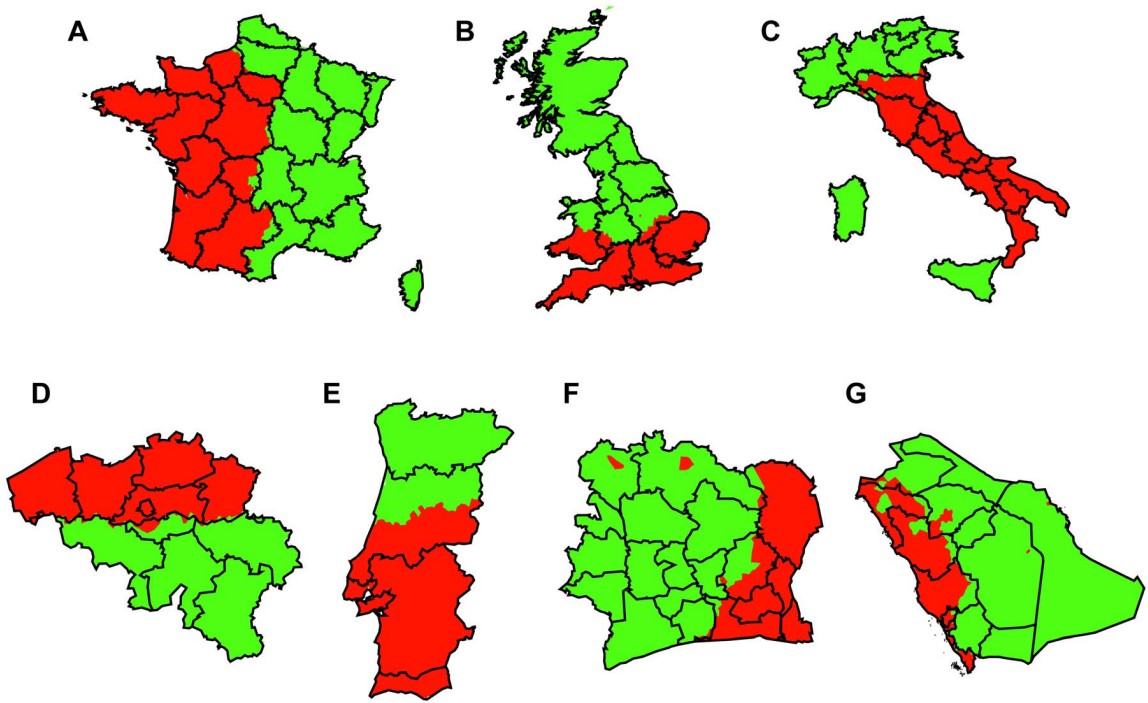

**Figure 4. Split of countries into two parts.** (A) France is split by a border going from center north to center south almost exactly following regional borders. (B) The UK splits along a west-east line which also splits Wales in two. (C) Mainland Italy is split along a line roughly following the northern border of Emilia-Romagna with the islands of Sardinia and Sicily being assigned to the northern part. (D) Belgium is split along the Dutch-French language barrier with Brussels assigned to the northern Dutch part. (E) Portugal is split roughly along the ancient border of the county of Portugal. (F) Ivory Coast and (G) Saudi Arabia are split into western and eastern parts.
doi:10.1371/journal.pone.0081707.g004

family members. Migration data would be needed to come to firmer conclusions.

The detected breaking line of Portugal, Fig. 4E, follows on the west side roughly the historical borders of the Condado Portucalense (county of Portugal), slightly south of the city of Coimbra and the Mondego river. This county of Portugal existed between the late ninth to the early twelfth century and was a fiercely disputed region between Moor and Christian reigns, with often shifting borders due to conquests and reconquests. This period marks the time in which the national identity of the Portuguese people was formed and the basis for the Portuguese kingdom was created. Given that the split is not very strong (a relatively large percentage of 12.1% of links exists between the split areas) it is not clear if the borderline we find can be reasonably attributed to these ancient regions or just to the surrounding areas of Lisbon and Porto. However, it is at least interesting to find a split into north and south as certain rivalries between those regions have left their imprints on almost every aspect of Portuguese social life [31].

These results do not come with any explicit policy implications due to the unclear causal relations, but the method can offer either careful attempts at historical insights into the evolution of specific communities, or provide possible "ground truth" to their cohesiveness if communication strength between the inhabitants is taken as a measure.

## Limitations and robustness

As the data sets under study feature call records that were provided by different sources, possibly collected and aggregated in different ways, a number of limitations and possible biases may influence the results. It is not possible to reliably and rigorously decide if and to which extent certain outcomes are caused by which reasons. For example, it is not clear why the partition of France corresponds much better to official regions than in other countries. Are there social or economic effects in place, where either individuals are more strongly separated by borders than in other countries, or where administrative regions have been defined in better agreement to existing social ties within the country? Or is it because of the high resolution of the data? One of the possible reasons is the attachment of the caller and callee to their actual locations taking into account both mobility and communication patterns. This also makes Portugal partitioning more clear compared to UK and Italy. However the case of highest fit for France probably results from a combination of all mentioned factors.

In any case, it is very difficult to separate such effects – additional detailed data on the long-time development of borders and social ties would be needed. However, we are able to at least assess the independence of the results from the used partitioning algorithm. For this task, we calculated partitions using the three additional, well-known processes of the Louvain method [21], the Clauset-Newman-Moore heuristic [32], and Newman's spectral division method [33]. Resulting $\mathcal{R}$ and $\mathcal{F}$ indices show no substantial deviations from our algorithm, Table S1 in File S1, asserting that the principle properties of the clusterings seem to be stable in terms of partitioning algorithm, while the quality of partition shapes slightly increases when higher modularity scores are obtained [23].

For a robustness analysis of the found partitions from fluctuations of the underlying networks, we performed a stability analysis where the networks were perturbed with various levels of random noise. Results are less clear, but show that community structures become unstable under high amounts of noise, while being somewhat stable under moderate noise, and are not affected at all by random fluctuations that appear in the execution of the algorithm itself, see File S1 and Figs. S1 and S2 in File S1.

It should always be clear that results of this work are based on communication data, and can possibly be subject to certain biases ranging from heterogeneous network coverage to data errors or quirks of the particular community detection algorithm used, see Fig. S3 in File S1. Due to the small number of approaches performed so far and the possible intricacies of the co-evolution of historical, socio-economic and political boundaries, it is not in every case clear if – in the case of conflicting outcomes – a result matching official boundaries should be interpreted as a validation of a method. However, future quantitative comparisons of different community detection methods, use of different data sets and a case by case treatment of specific regions should be able to resolve these issues. There already exists a vast corpus of literature on different methods of community detection other than modularity optimization [22], but it is an open question how exactly these different algorithms perform on spatial communication data, and if they are sensitive in terms of e.g. data resolution.

## Conclusion

Together with previous findings [10–13], we interpret our results as country-independent evidence for a general common pattern of human interactions in space, where landline and mobile phone call networks are taken as a proxy for interactions. The recurring properties we observe are threefold:

1. **Coherence**. The detected areas are coherent and almost never split up into disconnected components.
2. **Border-similarity**. Most of the boundaries of the detected areas closely follow existing political or socio-economic borders.
3. **Balance**. The number of detected areas takes a "natural", intermediate value between 1 and the number of nodes $N$, similar to the number of existing high-level administrative regions.

While the concept of coherence can be clearly defined and tested, and quantitative indices for border-similarity were provided, the case is not so obvious for "Balance". For a reasonable definition, apart from the similarity to the number of official regions, using high resolution data in future work it might be possible to involve and test classical theories of city spacing such as Central Place Theory [34].

Previous works [10–13] have demonstrated that the geographical projection of the community structure of human interaction networks in a few countries generates cohesive regions which generally follow the official regional boundaries. In the present work we validate this result for a number of differently scaled countries, using telephone calls as a proxy for human interactions. Further, we demonstrate that these properties are valid also for the smaller scale regional networks inside each of the detected major regions – the natural subpartitioning of the regional subnetworks again leads to geographically cohesive areas to many parts matching official borders of corresponding subregions. Finally, we show that some deviations might provide insights with possible practical implications in policy making. More work is however needed to systematically analyze the influence of population or cell tower distribution and heterogeneity on community detection, and longitudinal measurements on how fast and under which circumstances social interactions adapt to the introduction of new political borders.

In the present work we used both landline and mobile phone calls as a proxy for human interactions to uncover spatial regions of human connectedness and their agreements with or deviations from official regions. Results suggest that using the actual locations of customers (France, Portugal) leads to outcomes of higher quality. We expect future work on this issue to use various other types of human behavior such as commuting and travel data or economic activity for more refined insights in a "multiplex" view.

## Supporting Information

**File S1** Contains: **Table S1.** Partition differences to the administrative regions, for alternative algorithms. **Figure S1.** Cluster overlap index $\mathcal{R}$ comparing the noiseless partitions with partitions having different levels of noise, for Belgium and Portugal. **Figure S2.** Partitioning of Portugal with different levels of noise. **Figure S3.** Partitioning of Portugal with different algorithms.
(PDF)

## Acknowledgments

We thank Orange, British Telecom, Telecom Italia and Saudi Telecom Company for providing datasets for this research, Joseph K. Lee and Ritwik Yadav for GIS processing, and Kael Greco, Miriam Roure Parera and Hanna Lee for image processing. We further thank the National Science Foundation, the MIT SMART program, the Center for Complex Engineering Systems (CCES) at KACST and MIT, Audi Volkswagen, BBVA, The Coca Cola Company, Ericsson, Expo 2015, Ferrovial and all the members of the MIT Senseable City Lab Consortium for supporting the research.

## Author Contributions

Conceived and designed the experiments: SS MS RC TC ZS CR. Performed the experiments: SS MS RC TC ZS CR. Analyzed the data: SS MS RC TC ZS CR. Contributed reagents/materials/analysis tools: SS MS RC TC ZS CR. Wrote the paper: SS MS RC CR.